# Critical Separation of Clusters During Physical Vapor Deposition

L.G. Zhou and Hanchen Huang*

*Department of Mechanical Engineering, University of Connecticut, Storrs, CT 06269*

The critical separation of clusters, corresponding to the maximum density of clusters, affects growth characteristics during physical vapor deposition (PVD). In particular, this separation can affect surface smoothness in growing single-crystalline films, grain size distribution in growing polycrystalline films, and diameter in growing nanorods. This Letter reports a theoretical expression of the critical separation as a function of deposition conditions and accompanying verifications using lattice kinetic Monte Carlo (MC) simulations of PVD on Cu{111}. In contrast to existing theories, the theoretical expression in this Letter is (1) closed-form, (2) in better agreement with the MC simulations than the lattice approximation and (3) in better agreement with the MC simulations than the mean field approximation when the critical separation is large- larger than 25 nm for PVD on Cu{111}.

The separation of clusters on a surface or substrate during physical vapor deposition (PVD) changes with time [1]. At the start of deposition there are no clusters and their effective separation is large or infinitely large in theory. The density of clusters increases as deposition proceeds and reaches a maximum when their merging starts to dominate. Corresponding to the maximum density is the critical separation of clusters, which affects growth characteristics of thin films and nanorods.

The growth characteristics of single-crystalline films correlate with the critical separation but do not sensitively depend on it. Even if the critical separation is small, a layer-by-layer growth is still possible since small monolayer clusters can merge. Once the merging takes place the critical separation is no longer meaningful. For the case of polycrystalline thin films, their grain size depends on the critical separation; smaller critical separation results in smaller grain size. When it comes to the growth of nanorods, the critical separation is a controlling factor. In particular, the critical separation defines whether the growth of nanorods is feasible, and how large the diameter can be if it is feasible [2].

Theoretical studies of the critical separation date back to 1962 [3], and have since evolved in complexity and rigor [4-9]. In the following, we take a critical look at the existing theories, excluding numerical simulations such as Monte Carlo [10, 11] and level-set based approaches [12]. Conceptually, adatoms and clusters of atoms exist on a surface or substrate during deposition, and their concentrations change with time and spatial location. Due to the complexity of both spatial and time dependencies of these concentrations, a theoretical study usually makes approximations to focus on the key physics, leaving more rigorous solutions to direct numerical calculations. The two leading theoretical approaches are the lattice approximation [13] and the mean field approximation [6]. In order to identify the issues that require attention, we examine these two theories in sufficient detail. For clarity, without losing the principal features of these two theories, we consider only adatoms as mobile and all clusters as of equal size.

In the framework of lattice approximation, the adatom concentration $n(r,t)$ as a function of spatial location $r$ and time $t$ is governed by the following partial differential equation:

$$\frac{\partial n(r,t)}{\partial t} = \frac{\nu}{4}\nabla^2 n(r,t) + F \qquad (1)$$

This equation involves the adatom diffusion jump rate $\nu$ and deposition rate $F$, and is valid in an effective area surrounding a cluster. As a convention, length is in the unit of nearest neighbor distance, deposition rate is in the unit of monolayer (ML) per second, and adatom concentration is fractional. To solve this equation analytically, the quasi-steady state assumption is almost universally made, resulting in an explicitly time-independent equation for $n(r)$:

$$\frac{\nu}{4}\nabla^2 n(r) + F = 0 \qquad (2)$$

We note that condition of quasi-steady state is not always satisfied, and its guaranteed violation at the early stage of deposition is particularly problematic; we will discuss this problem again near the end of this Letter. Around each existing cluster (of circular shape), there is an effective circular area in which Equation (2) applies. The two boundary conditions of this equation in a circular area are that: (a) the flux is zero at the outer boundary of the circle, and (b) the adatom concentration is the equilibrium concentration, which is practically zero, at the inner circle that bounds the existing cluster. By relating the concentration gradient and concentration both as atomic flux at the inner circle, one defines the capture number of the cluster $\sigma_x$; and capture number of an adatom $\sigma_l$ in the same way. Through the capture numbers, the time dependencies of spatial-average adatom concentration $n_l$ and cluster concentration $N_x$ are governed by the following equations (do we need to mention the supersaturated condition that is implied here?):

$$\frac{dn_1}{dt} = F - 2\nu\sigma_1 n_1 n_1 - \nu\sigma_x n_1 N_x \quad (3)$$

$$\frac{dN_x}{dt} = \nu\sigma_1 n_1^2 - c(Z, N_x) \quad (4)$$

---

* Author to whom correspondence should be addressed; electronic mail: hanchen@uconn.edu

Where $c(Z, N_x) = N_x \left[ -1/Z + 4/(e^{4Z} - 1) \right] F$ is the rate of coalescence of clusters as a function of coverage Z [14]. The two drawbacks of the lattice approximation are that (a) the assumption of quasi-steady state is problematic – as further evidenced later in this Letter, and (b) the solution is not closed-form.

In comparison to the lattice approximation, the mean field approximation uses the same approach except the solution method of quasi-steady state adatom concentration in two ways. First, instead of an effective circular area around a cluster, the quasi-stead state equation is solved in an infinitely large area that contains homogenous sinks representing the clusters. This treatment effectively double counts the sink strength of a cluster. Second, the boundary condition at the outer circle is replaced by one at infinity, where the adatom concentration approaches its spatial-average value. The mean field approximation shares the same drawbacks as the lattice approximation, with subtle differences.

Having established the drawbacks of the existing theories, we use a conceptually different approach to achieve a closed-form expression of the critical separation. Further, we show that this expression is more reliable than the existing theories, or at least equally reliable in the case of small separations. Conceptually, we focus on the critical condition when the density of clusters is around its maximum, without tracking the time dependence of cluster size and density. In addition, we consider regularly patterned hexagonal regions, each containing a cluster of radius $R_i$, as shown in Fig. 1. Two nearby clusters are separated by a distance $L_s$. As a result of the regular pattern, the coalescence rate of clusters is zero. To facilitate an analytical solution of adatom concentration, we use a circular region in place of a hexagonal region and keep the area the same. The effective circular region has a radius $R_o$. Near the maximum density of clusters, or the critical separation, the probability of nucleating new clusters during deposition changes nearly abruptly with the separation. Based on this concept, we solve the quasi-steady state Equation (2) in the same way as in the lattice approximation to have:

$$n(r) = \frac{F}{\nu}\left( R_i^2 + 2R_o^2 \ln\frac{r}{R_i} - r^2 \right) \quad (5)$$

Instead of relating the spatial dependent adatom concentration $n(r)$ to its average, we use $n(r)$ directly in the determination of nucleation rate $\omega$ in the effective zone:

$$\omega = \int_{R_i}^{R_o} \nu n(r) n(r) 2\pi r dr$$

$$= \frac{\pi F^2 R_o^6}{6\nu}[7 - 30\frac{R_i^2}{R_o^2} + 15\frac{R_i^4}{R_o^4} - 2\frac{R_i^6}{R_o^6}$$

$$+ 124\frac{R_i^2}{R_o^2}\ln\frac{R_o}{R_i} + 24\left(\ln\frac{R_o}{R_i}\right)^2 - 36\ln\frac{R_o}{R_i}] \quad (6)$$

The total nucleation probability $P$ increases with time according to $dP/dt = \omega(1-P)$. In order to relate the inner radius $R_i$ with time, we first assume that there is no extra Ehrlich-Schwoebel (ES) barrier [15, 16], so: $R_i^2 = FR_o^2 t$ and

$$P = 1 - e^{-\int_0^{1/F}\omega dt} = 1 - e^{-\frac{7}{12}\frac{\pi F}{\nu}R_o^6} = 1 - e^{-(R_o/R_{oc})^6} \quad (7)$$

Where the critical outer radius is $R_{oc} = (12\nu/7\pi F)^{1/6}$. Or, in terms of the critical separation,

$$L_s = 2\sqrt{\pi/2\sqrt{3}}\left(\frac{12}{7\pi}\frac{\nu}{F}\right)^{1/6} \approx 1.9\left(\frac{12}{7\pi}\frac{\nu}{F}\right)^{1/6} \quad (8)$$

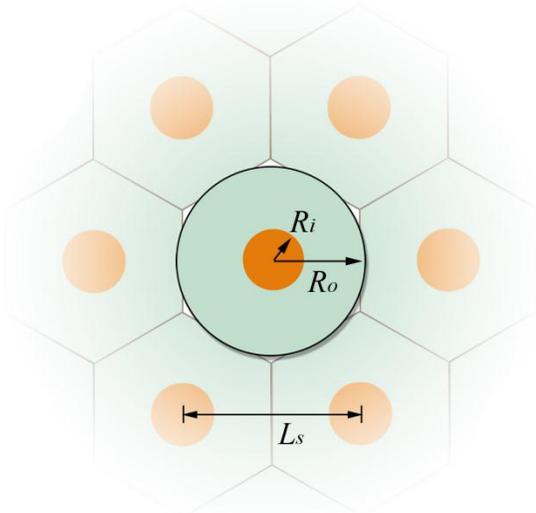

FIG 1 (color online). A regular pattern of hexagonal regions, each containing a cluster.

Next, we consider the effect of extra ES barrier. When the ES barrier is very large - say, infinitely large, $R_i^2 = R_o^2 - R_o^2 e^{-Ft}$, and the critical outer radius $R_{oc}$ will need to be obtained numerically. Our numerical results show that even an infinitely large ES barrier introduces less than 1% change to the critical outer radius $R_{oc}$, when $\nu/F$ is in the range of $10^7 \sim 10^{13}$, typical for PVD on Cu{111}.

In order to verify that the closed-form expression in Equation (8) is valid, we carry out lattice kinetic Monte Carlo simulations of PVD on Cu{111}. In the nearest neighbor model, each bond carries 3.54/12 (eV); the sublimation energy is 3.54 eV [17]. The diffusion barrier of an adatom on {111} surface is 0.06 eV according to *ab initio* calculations [18], and the prefactor of diffusion jump frequency is $0.5 \times 10^{12}$ s$^{-1}$ according to molecular dynamics simulations; as in our previous studies [19]. To ensure that clusters are compact, we also allow atoms having four or five neighbors to diffuse with a barrier of 0.25 eV, with the same prefactor of $0.5 \times 10^{12}$ s$^{-1}$. This value of 0.25 eV is consistent with *ab initio* calculations [20], and the results are insensitive to small variations of this value as long as clusters are compact. The substrate temperature is in the range of 200K to 520K, and the deposition rate $F$ is in the range of 0.05 to 512 ML per second. From each simulation, we identify the maximum density of clusters $N_x$, and derive the critical separation $L_s = 1.9/\sqrt{\pi N_x}$. The substrate area of the simulation cell is 1000nm x 1000nm. For each set of deposition conditions, we use 500 independent simulations to ensure that the standard deviation of $L_S$ is smaller than 0.5 lattice unit.

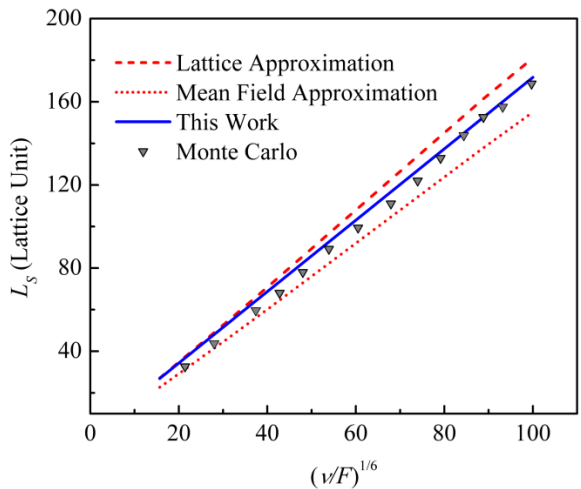

FIG 2 (color online). Comparison of theoretical expression in this work with lattice approximation and mean field approximation, in terms of the critical separation $L_S$ as a function of $\nu/F$; all with reference to lattice kinetic Monte Carlo simulations.

As shown in Fig. 2, the lattice kinetic Monte Carlo simulations verify our closed-form expression of the critical separation $L_S$, with minor discrepancy at small $L_S$. We note that our theory slightly overestimates the critical separation $L_S$ when it is small. It is known that a regular pattern of clusters – as is the case in our work – will absorb adatoms more effectively than a random distribution will [5]. As a result, theories based on a regular pattern of clusters overestimate the critical separation $L_S$. To appreciate why this overestimation causes more discrepancy at smaller critical separation, we formulate the lifetime of adatoms around a random distribution of clusters.

In this formulation, we represent the randomness by a variable circular area around each cluster. To simplify the formulation, we further assume that the existing cluster is small, its size is nearly zero, and that the outer radius of a circular area follows a uniform distribution between $R_o$-$\delta$ and $R_o$+$\delta$. For an area of radius $R$, the lifetime of an adatom before absorption by the cluster is $\pi R^2$ [8]. Averaging this lifetime over the uniform distribution of $R$, we have the average lifetime as $\pi \left( R_o^2 + \delta^2/3 \right)$. When $\delta$ is zero, this expression returns to the lifetime of adatoms around a regular pattern of clusters. According to this expression, deviation from a regular pattern of clusters results in longer lifetime of adatoms, and thereby higher nucleation rate and smaller critical separation. As $R_o$ increases, the contribution of $\delta$ to the lifetime decreases, and the discrepancy from different distributions also decreases. This analysis explains why the agreement between our theory and Monte Carlo simulations is better at larger critical separation.

Beyond verification of the closed-form expression, we further investigate whether and why the closed-form is more reliable than the lattice approximation and the mean field approximation. The comparison with the lattice approximation is more direct, since both approaches use the same quasi-steady state equation and the same boundary conditions. Fig. 2 shows that indeed the closed-form expression agrees with Monte Carlo simulations better than the lattice approximation does. To understand why this is the case, we track adatom concentrations using lattice kinetic Monte Carlo simulations at 300K and 32 ML/second. Fig. 3 shows the adatom concentration as a function of deposition time in the unit of coverage. The spatial-average adatom concentration from Equation (3) of the lattice approximation is comparable to that from lattice kinetic Monte Carlo simulations. However, the direct average of the spatial-dependent adatom concentration from Equation (2) of the lattice approximation is much higher than that from lattice kinetic Monte Carlo simulations. The inconsistency of these two averages of adatom concentrations in the lattice approximation is a result of unwarranted quasi-steady state assumption, and it results in erroneous nucleation rate at the early stage of deposition. This error is carried on as deposition continues. In contrast, our model does not have a history of cluster density and is less affected by the quasi-steady state assumption. In passing, we note that the mean field approximation has an intrinsic error of over counting the sink strength of clusters. Taking lattice approximation as reference, the mean field approximation over counts background sinks of clusters and thereby underestimates the flux into the reference clusters. As a result, the product of capture number and the spatial-average adatom concentration goes up, leading to higher nucleation rate and lower cluster separation.

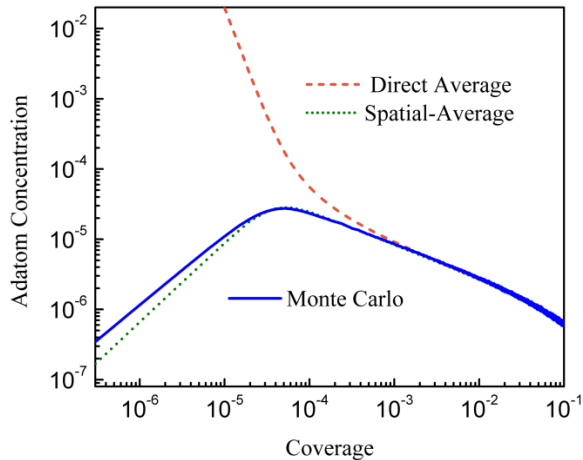

FIG 3 (color online). Inconsistency between the spatial-average adatom concentration and the direct average of spatial-dependent adatom concentration; all in reference to lattice kinetic Monte Carlo simulations.

Before closing, we briefly discuss how this theoretical work may help guide the synthesis of nanorods. According to the theory, the critical separation of Cu clusters can be on the order of 30 nm under typical PVD conditions. This value is not very sensitive to the change of adatom diffusion barrier due to the 1/6 power-law dependence; Equation (8). If Cu is deposited on a non-wetting substrate with the separation of clusters on the order of 30 nm, the clusters may grow vertically before coalescence. Under glancing angle deposition, the top of clusters receive more flux. The preferential flux on top and the non-wetting work together to promote the formation of multiple-layer surface steps [21], so as to allow the three-dimensional Ehrlich-Schwoebel barrier [18, 22] to operate. As a result, Cu nanorods of 30 nm in diameter may develop. This diameter is the smallest possible, since it is also the critical separation of clusters and the separation of clusters cannot be any smaller. According to our recent experiments [2], Cu nanorods of this small diameter are feasible, but could not be further reduced; an attempt to further reduction by increasing deposition flux led to a dense film.

In summary, we have developed a theoretical expression of the critical separation of clusters. This expression is closed-form, and is verified by accompanying lattice kinetic Monte Carlo simulations. In comparison with existing theories – the lattice approximation and the mean field approximation – this closed-form expression is in general more reliable.

**Acknowledgment:** The authors gratefully acknowledge the financial support of US Department of Energy (DE-FG02-04ER46167), and HH also acknowledges financial support of the National Science Foundation (CMMI-0856426) in bridging the growth of thin films with that of nanorods.

**References:**
[1] K. J. Routledge, and M. J. Stowell, Thin Solid Films **6**, 407 (1970).
[2] S. P. Stagon *et al.*, Appl. Phys. Lett. (2011). (submitted; arXiv:1112.2145v1).
[3] D. Walton, J. Chem. Phys. **37**, 2182 (1962).
[4] G. Zinsmeister, Thin Solid Films **2**, 497 (1968).
[5] R. M. Logan, Thin Solid Films **3**, 59 (1969).
[6] J. A. Venables, Philos. Mag. **27**, 697 (1973).
[7] J. A. Venables, G. D. T. Spiller, and M. Hanbucken, Rep. Prog. Phys. **47**, 399 (1984).
[8] J. Krug, Physica A **313**, 47 (2002).
[9] C. Ratsch, and J. A. Venables, J Vac. Sci. Tech. A **21**, S96 (2003).
[10] L.-H. Tang, J. Phys. I France **3**, 935 (1993).
[11] M. C. Bartelt, and J. W. Evans, Phys. Rev. B **54**, R17359 (1996).
[12] M. F. Gyure *et al.*, Phys. Rev. E **58**, R6927 (1998).
[13] M. J. Stowell, Philos. Mag. **26**, 349 (1972).
[14] R. Vincent, Proc. Roy. Soc. Ser. A **321**, 53 (1971).
[15] G. Ehrlich, and F. G. Hudda, J. Chem. Phys. **44**, 1039 (1966).
[16] R. L. Schwoebel, and E. J. Shipsey, J. Appl. Phys. **37**, 3682 (1966).
[17] C. J. Smith, *Metals Reference Book, 5th ed.* (Butter-worth, London, 1976).
[18] S. K. Xiang, and H. C. Huang, Appl. Phys. Lett. **92,** 101923 (2008).
[19] L. G. Zhou, and H. C. Huang, Phys. Rev. Lett. **101,** 266102 (2008).
[20] C. M. Chang, C. M. Wei, and S. P. Chen, Phys. Rev. Lett. **85**, 1044 (2000).
[21] R. X. Zhang, and H. C. Huang, Appl. Phys. Lett. **98**, 221903 (2011).
[22] S. J. Liu, H. C. Huang, and C. H. Woo, Appl. Phys. Lett. **80**, 3295 (2002).